\documentclass[aps,prb,showpacs,showkeys,twocolumn,groupedaddress]{revtex4-1}

\usepackage{graphicx, natbib}

\newcommand{\ee}[1]{\times 10^{#1}}

\newcommand{\dd}[3]{\frac{d^{#3} #1}{d #2^{#3}}}

\begin{document}

\title{Temperature dependence of the superheating field in niobium}

\author{N. R. A. Valles}
\email[]{nrv5@cornell.edu}
\author{M. U. Liepe}
\email[]{mul2@cornell.edu}

\affiliation{Cornell University, CLASSE, Ithaca NY 14853, USA}

\date{\today}

\pacs{74.25.-q, 74.25.Bt, 74.25.Ha}

\keywords{Niobium, Superheating Field, High Pulsed Power, Critical Magnetic Field, Material Studies, Quench Mapping, Oscillating Superleak Transducers, Ginsburg-Landau theory, Vortex Line Nucleation Model, Radio Frequency Superconductivity}

\begin{abstract}
This study experimentally investigates the temperature dependence of superheating field, $H_{sh}$, of niobium. Accurately determining this field is important both to test theory and to understand gradient limits in superconducting cavities for particle accelerators. This paper discusses theories that have been proposed in modeling the field and discriminates between them. The experimental procedure for measuring the temperature dependence of $H_{sh}$ utilizes high power pulses to drive a niobium cavity resonator, ramping up surface magnetic fields extremely quickly. The moment any part of the cavity transitions between the superconducting and normal conducting state can be determined by measuring the quality factor of the cavity as a function of time. Oscillating superleak transducers are used to demonstrate that the transition to the normal conducting state is global in nature, showing that a fundamental limit is encountered. Finally, we see that 110-120~$^\circ$C heat treatment of the cavity--a method commonly used to increase the quality factor at high accelerating gradients--may have the deleterious effect of reducing the superheating field of the material, which is the fundamental limiting factor in pursuing the maximal achievable accelerating gradient in superconducting niobium cavities.
\end{abstract}

\maketitle

\section{Introduction}

	An important property of niobium that has posed both theoretical and experimental challenges is the magnetic superheating  field. This field, also called the RF critical magnetic field, is the magnetic field at which a material undergoes a phase transition from the superconducting state to the normal conducting state.

	Superconductors can be broken up into two groups: those with positive surface energy and those with negative surface energy,  called Type~I and Type~II respectively. They can be quantitatively distinguished using the Ginsburg-Landau parameter, $\kappa_{GL}$, which is defined in terms of the London penetration depth, $\lambda_L$ and the coherence length, $\xi_0$ of the material according to $\kappa_{GL} = \frac{\lambda_L}{\xi_0}$.\cite{Cyrot} A superconductor is Type~I if $\kappa_{GL}<1/\sqrt{2}$ and Type~II otherwise. Further discussion of these implications is presented in the theory section of this paper.

	Type~I superconductors exclude DC magnetic fields up to a lower critical field $H_c$, after which it becomes energetically favorable for magnetic flux to penetrate the material and cause a transition to the normal conducting state. Type~II superconductors also exclude flux at low DC magnetic fields, but above the lower critical field $H_{c_1}$ magnetic flux penetrates the material, creating an array of normal conducting and superconducting regions. As the field is increased, the surface area of the normal conducting regions with additional flux entry increases until the entire sample is normal conducting at a field of $H_{c_2}$.

	For both Type~I and II superconductors, a metastable state exists for fields above their lower critical fields, where flux penetration is delayed, and the material remains fully superconducting in the Meissner state.\cite{PhysRevLett.12.14, Sethna}
	
	The semiclassical approach for solving the superheating field  requires solving the Eilenberger equations.\cite{Eilenberger} This approach has been taken to solve for the temperature dependence of the superheating field for very strong Type~II materials.\cite{Sethna} Niobium is not a high-$\kappa$ material, so thoroughly solving for $H_{sh}$ is more challenging.

	 An approximate solution of the superheating field is the Ginsburg-Landau theory which predicts the superheating field goes as
\begin{equation}
	H_{sh}(T) = c(0) H_c \left[ 1 - \left(\frac{T}{T_c}\right)^2 \right],
	\label{eq:GLTheory}
\end{equation}
where for niobium, the thermodynamic critical field $H_c = 2000$~Oe, and the critical temperature $T_c = 9.2$~K. The constant $c(0)$ is the ratio of the superheating field and the thermodynamic critical field at zero temperature.~\cite{Dolgert} For very high purity niobium, $c(0)\approx1.2$.\cite{RFSuper} Note that this phenomenological approach is only valid for $T_c-T\ll T_c$.

	Another theory put forward by K. Saito and T. Yogi is the Vortex Line Nucleation Model.\cite{Saito, Yogi} This theory proposes that the superheating field varies with temperature according to
\begin{equation}
	H_{sh} = H(0) \left[1 - \left(\frac{T}{\bar{T}_c}\right)^4 \right],
	\label{eq:VLNModel}
\end{equation}
where $H(0) = 1780.4$~Oe and $\bar{T}_c = 9.014$~K is the reduced critical temperature of the material. It is important to note that both theories are consistent with previous experiments near $T_c$ but differ from each other significantly at lower temperatures.

\begin{figure}[htbp]
	\centering
		\includegraphics[width=85mm]{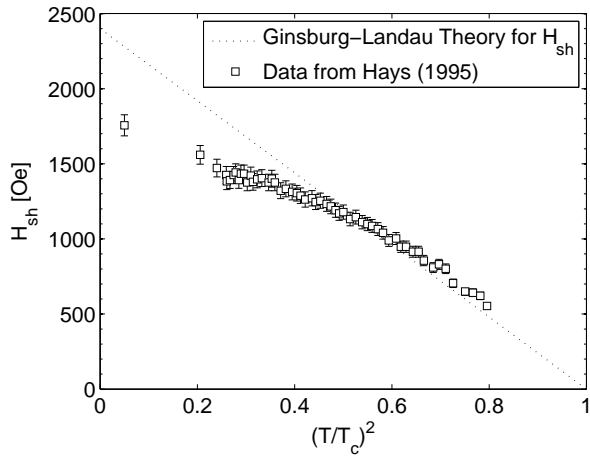}
	\caption{Measurement of the superheating field of niobium from Hays and Padamsee, using a 1.3~GHz cavity that received a buffered chemical polish surface treatment and did not receive a final low temperature heat treatment.\cite{HaysHPP} Ginsburg-Landau theory predicts $H_{sh} = 1.2 H_c [1-(T/T_c)^2]$ near $T_c$ for pure niobium. At high temperatures, the data agrees with theory, but flattens out at low temperatures.}
	\label{fig:HaysData}
\end{figure}

	Aside from the theoretical challenges, measuring the superheating field has also posed experimental challenges. Measurements of $H_{sh}(T)$ have been carried out for niobium, lead and Nb$_3$Sn.\cite{Campisi87, HaysHPP, NickPAC} Both Campisi and Hays show that near $T_c$, $H_{sh}$ is linear with $(T/T_c)^2$ with a slope of $\sim 1.2 H_c$~Oe, but far from $T_c$, the maximum achieved RF critical field only weakly depends on temperature. Hays' data is shown in Fig.~\ref{fig:HaysData}.
	
	The reduced slope of the maximum field for $\frac{T}{T_c} \ll 1$ may be caused by effects such as heating by an increase in surface resistivity, electron field emission, or normal conducting defects. Any of these effects may have suppressed the results from reaching the experimental superheating field at these temperatures, so further work should be done to insure that a fundamental limit is reached.

	Finally, the accurate determination of the superheating field also is of interest in application. While theory and experiment agree near $T_c$, niobium microwave cavities for particle accelerators operate at temperatures $<T_c/4$, where theory and experiment are disparate. Measurements can set stringent upper bounds for what maximum accelerating gradients are theoretically achievable in these machines.

\section{Theory}

	Niobium is a Type~II superconductor, meaning it has a negative surface energy. One of the first phenomenological attempts to address superconductors with negative surface energy was Ginsburg and Landau's treatment.\cite{GLTheory} They found that the superheating field near $T_c$ has the form described in Eq. \ref{eq:GLTheory}.

\begin{figure}[htbp]
	\centering
		\includegraphics[width=0.50\textwidth]{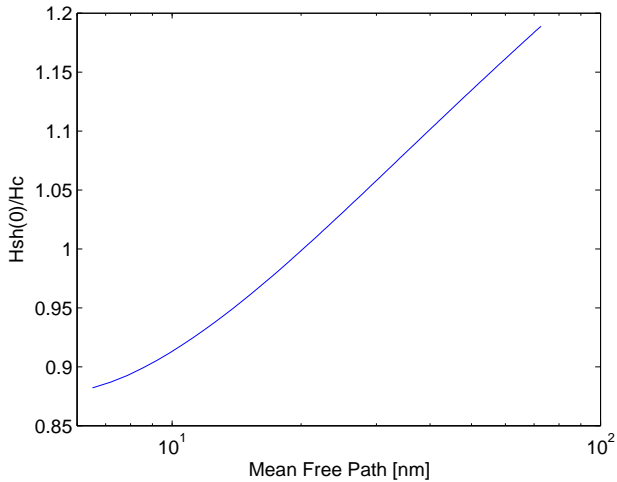}
	\caption{Plot of the ratio of the superheating field divided by $H_c$ at zero temperatures versus the electron mean free path for niobium.}
	\label{fig:HshHcVsL}
\end{figure}

	The constant $c(0)\equiv H_{sh}(0)/H_c$ in Eq.~\ref{eq:GLTheory} is dependent on the order parameter of the material, $\kappa = \lambda/\xi$, which is the ratio of the penetration depth and coherence length of the material. The Eilenberger equations\cite{Sethna} have been solved near $T_c$ to yield Fig.~\ref{fig:HshHcVsL}.\cite{Transtrum}

\begin{figure}[htbp]
	\centering
	\includegraphics[width=0.50\textwidth]{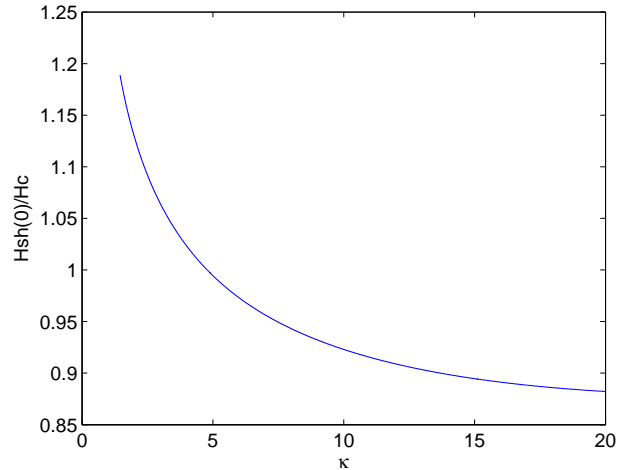}
	\caption{Plot of the ratio of the superheating field and the thermodynamic critical field, $H_c$, at zero temperature, versus the Ginsburg-Landau parameter, $\kappa$.\cite{Transtrum}}
	\label{fig:HshHcVsKappa}
\end{figure}

	Very pure samples of niobium have electron mean free paths, $\ell$, that are large compared to the coherence length and London penetration depth. As the purity of the niobium decreases, it enters the `dirty' regime where $\ell \ll \lambda (\sim \xi)$. Impurities can depress the critical temperature of the superconductor, which in turn affects the field $H_c$ such that $T_c \propto H_c$.~\cite{RFSuper}

	Pure niobium has a coherence length of $\xi_0 = 64$~nm, and a London penetration depth $\lambda_L = 36$~nm.\cite{NbProps} The coherence length, $\xi$ and penetration depth, $\lambda$ of a sample with a given mean free path, $\ell$ can be calculated--near absolute zero--according to Pipard's equations:\cite{Pipard}
\begin{eqnarray}
	\frac{1}{\xi} &=& \frac{1}{\xi_0} + \frac{1}{\ell},\\
	\lambda &=& \lambda_L \sqrt{1 + \frac{\xi_0}{\ell}}.
\end{eqnarray}
It is straightforward to write the Ginsburg-Landau parameter as a function of $\ell$:
\begin{equation}
	\kappa(\ell) = \frac{\lambda_L}{\xi_0} \left(\frac{\xi_0+\ell}{\ell}\right)^{3/2}
	\label{eq:KappaOfL}
\end{equation}
Using Eq.~\ref{eq:KappaOfL} the superheating field coefficient, $c(0)$, can be plotted versus the electron mean free path, shown in Fig.~\ref{fig:HshHcVsL}.

\section{Experimental Method}

	To measure niobium's superheating field, a 1.3~GHz resonating cavity was driven by short (200~$\mu$s) high power (up to 1.5~MW) pulses from a klystron. These pulses raise the fields in the cavity to their maximum values over $\sim100$~$\mu$s. Increasing the fields in this manner prevents the cavity from heating significantly, meaning that the temperature measured at the outer cavity surface is also the temperature of the inner surface. When the magnetic field on the cavity surface reaches the superheating field, $H_{sh}$, the niobium wall undergoes a phase transition into the normal conducting state.

\subsection{Calculation of $Q_0$ as a function of time}

	To accurately measure the superheating field, it is essential to determine precisely when the cavity transitions to the normal conducting state. Previous work has shown that a niobium cavity remains at least 90\% superconducting as long as the intrinsic quality factor is greater than $2 \ee{6}$.\cite{HaysHPP, HaysQ} It has been shown how to determine the quality factor as a function of time;\cite{Farkas84, Campisi84, HaysHPP, HaysQ} we merely reproduce the argument here for completeness.

	A cavity driven on resonance, $\omega$, by a single input coupler with an incident power, $P_f$, reflects some power, $P_r$, stores energy in the field, $U$, and dissipates some energy in the cavity walls, $P_{d}=\frac{\omega U}{Q_0}$. Conservation of energy gives:
\begin{equation}
	P_f = P_r + \frac{\omega U}{Q_0} + \dd{U}{t}{}
	\label{eq:ConservationOfEnergy}
\end{equation}

	The reflected power is not a measured quantity, so another expression relating $P_r$ and $P_f$ is needed. A full derivation of the needed equation is presented in Padamsee et. al., Chap. 8;\cite{HasanEq} in this paper only a plausibility argument will be made, and the result quoted.

	The reflected power is the superposition of the reflection of the incident power signal, and the power emitted from the cavity through the coupler. The reflection coefficient of the cavity can be expressed in terms of the admittances of the waveguide and the cavity-coupler system. Expressing these admittances in terms of cavity parameters one finds
\begin{equation}
	\label{eq:ReflectedPower}
	\sqrt{P_r} = \sqrt{P_f} - \sqrt{P_e}.
\end{equation}
where $P_e = \frac{\omega U}{Q_{ext}}$ is the power losses through the coupler with an external quality factor of $Q_{ext}$.

	Using $P_r$ from Eq.~\ref{eq:ReflectedPower} in Eq.~\ref{eq:ConservationOfEnergy} yields the expression
\begin{equation}
	\frac{\omega U}{Q_0} = 2 \sqrt{\frac{\omega U P_f}{Q_{ext}}} - \dd{U}{t}{} - \frac{\omega U}{Q_{ext}}
\end{equation}
The final expression can be obtained by using the identity $\dd{\sqrt{U}}{t}{}=\frac{1}{2\sqrt{U}}\dd{U}{t}{}$ to yield
\begin{equation}
\frac{1}{Q_0} = \frac{2}{\omega \sqrt{U}}\left(\sqrt{\frac{\omega P_f}{Q_{ext}}} - \dd{\sqrt{U}}{t}{}\right) - \frac{1}{Q_{ext}}
\label{eq:QvsTime}
\end{equation}

	Equation~\ref{eq:QvsTime} allows one to calculate $Q_0$ as a function of time from measurements of $P_f$ and $U$. Finding the time when the quality factor of the cavity falls bellow $2 \ee{6}$ pinpoints when the cavity transitions into the normal conducting state.\cite{HaysHPP}

\subsection{Quench characterization using second sound}

There are several mechanisms that can initiate a phase transition from the superconducting to the normal conducting state. Defects in the cavity surface can cause heating, leading to thermal breakdown and quench. Field emission, multipacting and dust in the cavity can also be sources of quench due to heating. To ensure that the quench is caused by the cavity making a global transition to the normal conducting state when the superheating field is exceeded, oscillating superleak transducers (OSTs) were used to determine where the observed cavity quenches originate.

	Energy dissipated from the outer cavity surface couples to a second sound wave in the superfluid Helium surrounding the cavity. The OSTs measure the arrival time of the second sound wave traveling through superfluid Helium, and by using multiple transducers, the origin of the second sound wave can be triangulated.\cite{Zac}

	A quench caused by a defect on the surface of the niobium will cause the field energy to be dissipated from a point like source. On the other hand, a transition caused by exceeding the superheating field of the cavity will be global, and the second sound wave should originate from the entire high magnetic field region of the cavity simultaneously. The second sound wave measurement thus allows us to determine whether the transition is caused by a local defect or is initiated by exceeding the superheating field.

\subsection{Cavity details and experimental setup}

	 The cavity used in the experiment is a 1.3~GHz single-cell re-entrant design.\cite{VSDesign} It was constructed of high RRR niobium with a bulk RRR of approximately 500, which corresponds to a mean free path of $\sim1800$~nm.
	 
	 The cavity's history prior to this experiment has been detailed elsewhere.\cite{NickSRF} Before this experiment, the cavity received a vertical electropolish,\cite{VEP} which uses a mixture of HF and H$_2$SO$_4$ to remove material from the inner surface. The electropolish removed 5-10~$\mu$m of material  while the electrolyte temperature was maintained below 20~$^\circ$C. It was high pressure rinsed for 2~hours, mounted on a test stand and evacuated. While under vacuum, the cavity was surrounded in 110~$^\circ$C air for 48~hours, a process that can prevent the quality factor from deteriorating at high fields.\cite{Eremeev}

\begin{figure}[htb]
    \centering
    \includegraphics*[width=85mm]{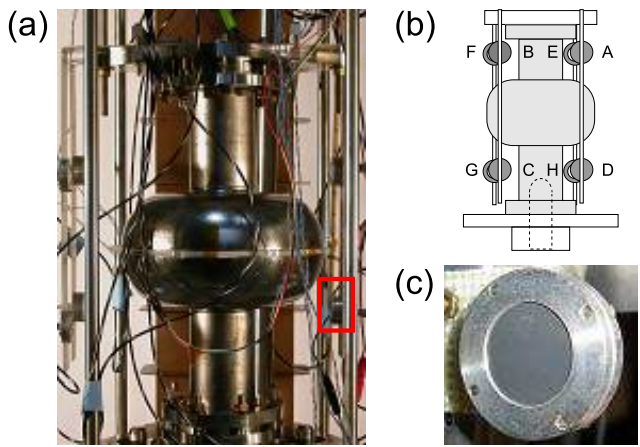}
    \caption{(Color online) (a) Experimental set-up of 1.3~GHz re-entrant cavity mounted on a test stand. The copper waveguide behind the cavity connects to the klystron which supplies the high pulsed RF power. Eight OSTs are mounted at corners of a cube around the cavity and are used to detect quench locations. A red box bounds a bottom right OST. (b) Side-view schematic of experimental set-up. OSTs labeled A-D are in near plane, and F-H are in far plane. (c) Front view of an OST. The dark gray membrane is semipermeable, allowing superfluid helium to penetrate, which changes the capacitance of the OST.}
    \label{fig:setup}
\end{figure}

	Three Cernox temperature sensors were mounted on the outside of the cavity to measure niobium wall temperature, and a Germanium thermometer measured the bath temperature. Eight OSTs were placed roughly at the points of a cube surrounding the cavity. A photograph of the experimental setup is shown in Fig.~\ref{fig:setup}. The entire setup was placed in a liquid Helium bath, which was used to control the temperature of the cavity.

\section{Results}

\subsection{Continuous wave results}

\begin{figure}[htbp]
	\centering
		\includegraphics[width=0.50\textwidth]{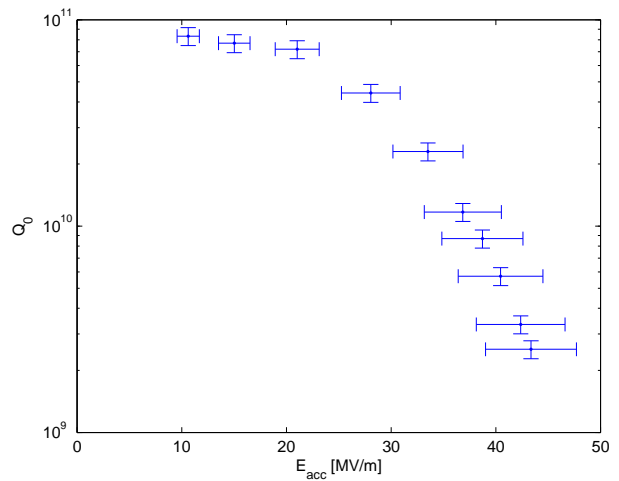}
	\caption{A continuous wave Q vs E curve taken at 1.6 K for cavity LR1-3. The Q degrades as the accelerating gradient is increased.}
	\label{fig:QvsE_1p6}
\end{figure}

	The cavity was first tested in continuous wave (CW) mode to measure its properties. Its intrinsic quality factor as a function of accelerating gradient is shown in Fig.~\ref{fig:QvsE_1p6}. The cavity reached accelerating electric fields up to 42~MV/m in CW, corresponding to a maximum surface magnetic field of 1474~Oe, and demonstrates a strong decrease in $Q_0$ (i.e. increase in surface resistivity) at high fields. The reduced quality factor at high accelerating gradients is discussed in Sec.~\ref{discussion}. In spite of the degradation of the quality factor at high fields, these heating losses do not cause significant global heating in pulse mode cavity operation.

\begin{figure}[htbp]
	\centering
		\includegraphics[width=0.50\textwidth]{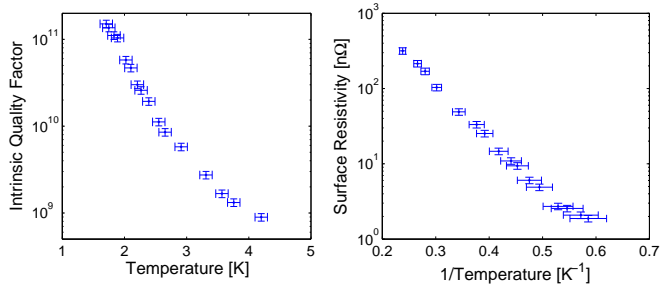}
	\caption{Left: Quality factor of the cavity as a function of temperature. The cavity had a quality factor of $1.5\ee{11}$ at 1.7~K. Right: The Surface resistivity of the cavity as a function of temperature. The residual resistivity of the cavity is $(0.92 \pm 0.23)$~n$\Omega$. Data was taken at an accelerating gradient of 6~MV/m.}
	\label{fig:QandRsVsTemp}
\end{figure}

	The quality factor was also measured as a function of temperature in CW. This information can be used to determine the surface resistivity of the cavity from the relation $R_s = G/Q_0$, where $G$ is the geometry factor (283.1~$\Omega$ for the cavity tested) and $Q_0$ is the intrinsic quality factor. Plots of these quantities versus temperature are presented in Fig.~\ref{fig:QandRsVsTemp}.

	The surface resistivity of the cavity has two contributions, a temperature independent residual resistivity, $R_0$, and the BCS surface resistivity:
\begin{equation}
	R_s(T) = R_0 + A(f,\ell,T)\exp\left(-\frac{\Delta}{k_b T_c}\frac{T_c}{T}\right)
\end{equation}
where $A$ is a function of the frequency, $f$, mean free path, $\ell$ and temperature, $T$. The residual resistivity of the cavity is found to be $R_0 = (0.92 \pm 0.23)$~n$\Omega$ from the low temperature data.

	Subtracting the residual resistivity from the surface resistivity leaves the BCS resistivity which can be calculated from material properties. A Fortran code, SRIMP,\cite{Halbritter} based on the Halbritter definitions,  was used to fit the measured BCS surface resistivity by varying the energy gap, $\Delta(0)$ and mean free path, $\ell$, of the material. The parameters used in the data fit are listed in Table~\ref{table:SRIMPParams}.
\begin{table}[h]
	\begin{center}
		\begin{tabular}{ll}
		\hline
		Parameter & Value\\
		\hline
		Frequency & 1294.5~MHz\\
		Critical Temperature  & 8.83~K\\
		Coherence Length & 640 \AA\\
		London Penetration Depth & 360 \AA\\
		\hline
		\end{tabular}
		\caption{\label{table:SRIMPParams} Material properties used in the calculation of BCS resistivity of the niobium. The critical temperature of the niobium is depressed, as discussed in Sec. \ref{sec:PulsedMeasurements}.}
	\end{center}
	\vspace{-0.6cm}
\end{table}

\begin{figure}[htbp]
	\centering
		\includegraphics[width=0.50\textwidth]{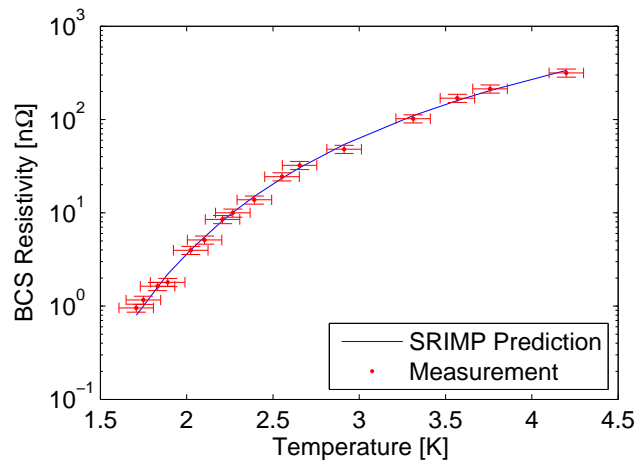}
	\caption{The BCS resistivity versus temperature. The blue line is the results of a code, SRIMP, which calculates BCS resistivity from material properties. This fit gives the mean free path estimate of $\ell = 26.91 \pm 1.19$~nm.}
	\label{fig:BCSvsTemp}
\end{figure}

	The resulting fit is displayed in Fig.~\ref{fig:BCSvsTemp}, and was fit with parameters $\Delta(0)/(k_B T_c)=2.14 \pm 0.03$ and $\ell = 26.91 \pm 1.19$~nm. This small mean free path is consistent with results after baking obtained by Ciovati.\cite{Ciovati} From Eq.~\ref{eq:KappaOfL} the mean free path corresponds to a Ginsburg-Landau parameter $\kappa = 3.49 \pm 0.16$, showing that the decreased mean free path of the niobium causes it to become a more strongly Type~II superconductor. Referencing Figs.~\ref{fig:HshHcVsKappa} yields $H_{sh}(0)/H_c = 1.044 \pm 0.001$. For the cavity tested the Ginsburg-Landau prediction for the superheating field is then
\begin{equation}
	H_{sh}(T) = (1.044 \pm 0.001) H_c \left[ 1 - \left(\frac{T}{T_c}\right)^2\right].
	\label{eq:GLPred}
\end{equation}

\subsection{\label{sec:PulsedMeasurements}Pulsed measurements}

\begin{figure}[htbp]
	\centering
		\includegraphics[width=0.50\textwidth]{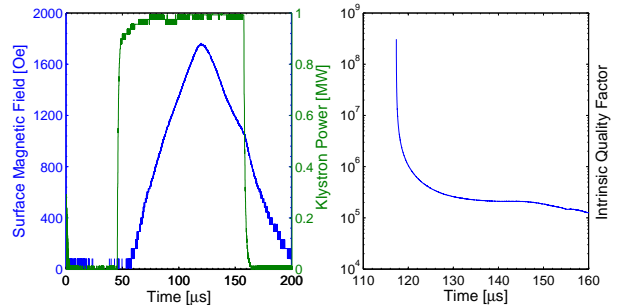}
	\caption{(Color online) Left: Trace showing the square power wave incident on the cavity (green) and the surface magnetic field of the cavity (blue) taken at 3.0~K. The pulse is sharply peaked around $\sim120\mu$s. Right: Plot of $Q_0$ versus time. The phase change occurs at 118~$\mu$s, which corresponds to $H_{sh} = 1749$~Oe.}
	\label{fig:Trace2p96K}
\end{figure}

\begin{figure}[htbp]
	\centering
		\includegraphics[width=0.50\textwidth]{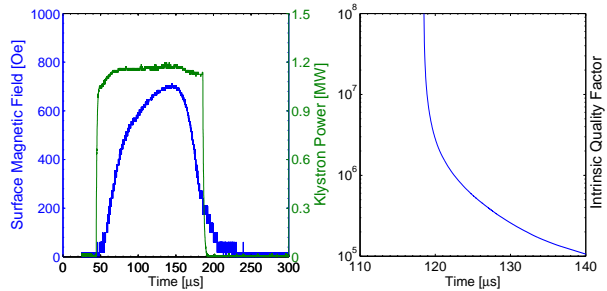}
	\caption{(Color online) Left: A trace showing the square power wave incident on the cavity (green) and the surface magnetic field of the cavity (blue) taken at 7.2~K. Right: Plot of $Q_0$ vs time. Compared to the 3.0~K case, the peak is broadened, so accurately determining time of phase transition is essential. The plot shows that $Q = 2\ee{6}$ at $120.6$~$\mu$s. This corresponds to $H_{sh}= 664$~Oe.}
	\label{fig:Trace7p2}
\end{figure}

	A klystron was was used to drive the cavity on resonance with high power pulses. The incident power and field in the cavity during two pulses are at different temperatures are shown in Figs.~\ref{fig:Trace2p96K} and \ref{fig:Trace7p2}. The phase transition occurs when the intrinsic quality factor of the niobium reaches $2\ee{6}$. At low temperatures the superheating field is approximately given by the peak value of the magnetic field. At higher temperatures, however, the field in the cavity is not as sharply peaked, instead showing a broadened structure. In this case it is imperative to have a accurately determine of $Q_0$ as a function of time to pinpoint when the phase transition occurs.

	A transition to the normal conducting state can be caused by defects in the material or field emission leading to heating--among other things. To obtain further certainty that the quench is due to reaching the fundamental limit of the superheating field, OST measurements were used to determine the origination of the heating.

\begin{figure}[htbp]
   \centering
   \includegraphics[width=0.50\textwidth]{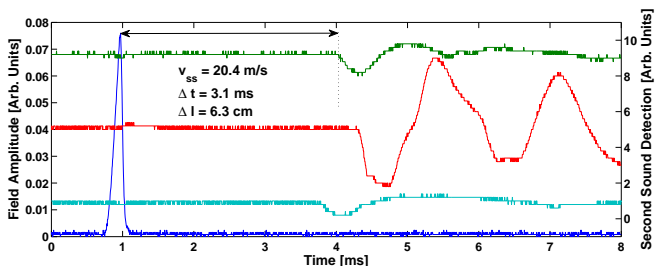}
   \caption{OST data for a single quench event. The lowest trace is the field in the cavity, and the three upper traces are 3 of the 8 OST signals, corresponding to sensors A-C as denoted in Fig. \ref{fig:setup}, all in arbitrary units. The spike at 1~ms is the cavity filling and emptying. The oscillations 3~ms later is the detection of second sound waves by the OSTs. The small time discrepancy between when the OSTs begin to oscillate correspond to the small differences between OST distances from the high magnetic field region of the cavity. Though not shown,  all 8 OSTs trigger simultaneously, demonstrating the cavity quench is a global event.}
   \label{fig:GlobalQuenchEvent}
\end{figure}

	A quench event as recorded by OSTs is shown in Fig.~\ref{fig:GlobalQuenchEvent}. All eight OSTs registered a signal a few milliseconds after each quench event. The energy dissipated in the cavity travels through the bulk of the material and at the outer surface of the cavity couples to a second sound wave in the liquid Helium, which is detected by the OST array.

	Measurements show that the observed triggering signature of the OSTs only occurs when the entire high magnetic field region quenches simultaneously. This shows that the entire high magnetic field portion of the cavity transitioned to the normal conducting state at the same time, characteristic of reaching a fundamental limit as opposed to a local effect.

	Each of the quenches measured below the lambda point of Helium, (Helium must be superfluid for a second sound wave to travel), were found to be global in origin, meaning that each OST measured the source of the second sound wave as originating from the nearest point of the cavities high magnetic field region. This adds credence to the assertion that a fundamental quantity was measured.

\begin{figure}[htbp]
	\centering
		\includegraphics[width=0.50\textwidth]{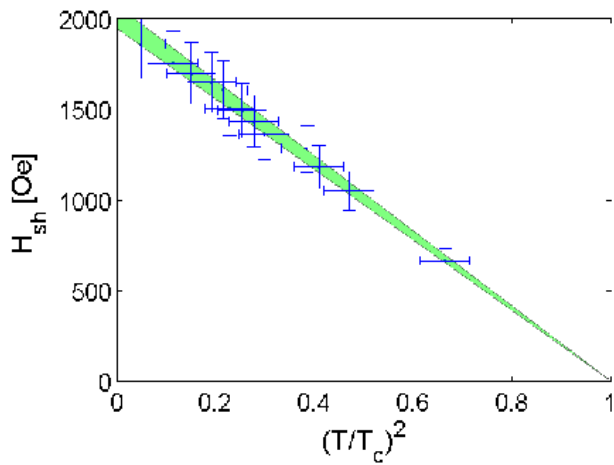}
	\caption{(Color online) Measurements of the superheating field plotted versus $(T/T_c)^2$, where $T_c=8.83$~K. The green cone is the Ginsburg-Landau prediction for niobium with a mean free path of $26.91\pm1.19$~nm, including the uncertainty in the ratio of $H_{sh}(0)/H_c$.}
	\label{fig:FinalHshvsT}
\end{figure}

	The pulsed measurements of the superheating field are presented in Fig.~\ref{fig:FinalHshvsT}. A linear fit was performed on the data, giving the result
\begin{equation}
	H_{sh}(T) = \left[(1979 \pm 58) - (2151 \pm 130) \left(\frac{T}{T_c}\right)^2\right] \text{Oe}
	\label{eq:MeasurementFit}
\end{equation}
where $T_c = 9.22$~K for pure niobium.

By definition, the superheating field should vanish at the critical temperature. Our data suggests that the cavity's critical temperature has been depressed due to impurities in the material. The data gives the sample's critical temperature as $\tilde{T}_c = 8.83 \pm 0.29$~K, which corresponds to a depressed critical field of $H_c = (1920 \pm 64)$~Oe, since $H_c \propto T_c$, and the clean value of $H_c$ is 2000~Oe. Inserting this value for the depressed critical field into Eq.~\ref{eq:GLPred} yields the prediction for the superheating field for $\kappa=3.5$ of
\begin{equation}
	H_{sh}(T) = (2003 \pm 64)\text{ Oe }\left[1-\left(\frac{T}{\tilde{T}_c}\right)^2\right].
	\label{eq:FinalPred}
\end{equation}	
 This prediction is plotted with the data in Fig.~\ref{fig:FinalHshvsT} and is consistent with the measured values of the superheating field.

\section{\label{discussion}Discussion}

	This experiment yielded several remarkable results. First, the surface preparation of vertical electropolishing the cavity produced a niobium cavity with a very high quality factor,  measured at 1.7~K, of $1.5 \times 10^{11}$ at 6~MV/m. This corresponds to a residual resistivity of $(0.92 \pm 0.23)$~n$\Omega$ which is among the smallest measured. 	The BCS resistivity of the cavity is also low, with values of 1.95~n$\Omega$ at 2.0~K ($Q_0 = 5.8\ee{10}$) and $3.16\ee{2}$~n$\Omega$ at 4.2~K ($Q_0 = 8.9\ee8$). Niobium has minimal BCS resistivity for mean free paths between 10-50~nm.\cite{Kneisel} The obtained mean free path of 26~nm, is within 5\% of the minimum value, showing that the 100-120~$^\circ$C heating process is necessary for achieving very low surface resistivities. Reliably producing cavities at this performance level is essential for next generation CW superconducting accelerators such as Project X,\cite{ProjX} and linac driven light sources such as Cornell's ERL.\cite{ERL} For CW machines, very small surface resistivities are more important than highest accelerating gradients because of the large cryogenic load in continuous cavity operation.

	As shown in Fig.~\ref{fig:QvsE_1p6}, the quality factor of the cavity significantly degraded as the accelerating gradient increased. Generally, the 110~$^\circ$C heat treatment procedure is known to reduce the high field Q-slope. However, previous work has shown that the effectiveness of this procedure varies, even for electropolished cavities.\cite{Gigi2007} The high field Q-slope is emphasized by the fact that the cavity has a very high initial $Q_0$ at low fields, and reaches gradients above 40~MV/m.

	High RRR materials are commonly used in superconducting RF cavities to improve thermal conduction. This experiment used a cavity with a bulk RRR of $\sim500$. However the RF properties of the cavity are determined by the properties of the surface layers with a thickness of $\sim50$~nm. The 110~$^\circ$C heating process reduced the mean free path of the surface layer to about 26~nm, corresponding to a RRR of only $\sim4.5$.

	Baking a cavity at $110-125$~$^\circ$C is currently part of routine cavity preparation, since it is the only known method to minimize the undesirable high field $Q$-slope at high accelerating gradients.  However, by reducing the mean free path, this process has the undesirable effect of increasing the Ginsburg-Landau parameter, thereby significantly reducing $H_{sh}$ by as much as 25\% (see Fig.~\ref{fig:HshHcVsL}). As the data near $T_c$ shows in Fig~\ref{fig:HaysData}, a slope of $1.2H_c$ is possible if the niobium does not recieve a low temperature heat treatment. This is a 20\% increase from value measured for the 110~$^\circ$C heat treated cavity in this study. Thus, for applications requiring high quality factor cavities at ultimate accelerating gradients, it is necessary to find a process other than baking that removes the high field Q-slope without reducing the mean free path of the electrons.

	OST data verified that the quenches were globally initiated, consistent with reaching the superheating field. The quenches were not locally initiated, eliminating point defect heating, field emission or contamination from limiting the maximum magnetic fields.

	The experimental data for the superheating field of niobium agrees with the Ginsburg-Landau prediction of $H_{sh}$ being linear in $(T/T_c)^2$ down to 1.7~K within the $\pm10\%$ measurement uncertainty. This is a result that has never before been obtained. Moreover, this theory correctly predicts an accurate slope for the $(T/T_c)^2$ dependence for niobium once the reduction in the mean free path in the RF surface layer by the cavity preparation has been taken into account. This evidence, along with the fact that superheating fields as large as $\sim$1900~Oe have been measured, excludes Saito's theory from modeling the superheating field.

\section{Conclusions}

	The superheating field of niobium was shown to be linear versus $(T/T_c)^2$, in accordance with Ginsburg-Landau theory, down to a temperature of 1.7~K. State of the art surface treatments have resulted in a cavity with a high quality factor and very low residual resistivity. The low temperature baking process has been shown to adversely effect the superheating field of niobium by decreasing the mean free path of the RF surface layer, which is an impediment to obtaining the ultimate achievable accelerating gradients in niobium cavities for next generation particle accelerators.

\section{Acknowledgments}

The authors thank J. Sethna and M. Transtrum for their theoretical calculations of $H_{sh}(0)/H_c$ versus the Ginsburg-Landau parameter, Z. Conway for help with OST instrumentation and vertical electropolishing the cavity and S. Posen for assistance during the cavity test. They also thank the Alfred P. Sloan Foundation and the Department of Energy under contract DE-SC00002329 for funding this research.

\bibliographystyle{apsrmp4-1}
\bibliography{./SFoNBib}

\end{document}